\documentclass[twocolumn,showpacs,prl]{revtex4} 
\usepackage{graphicx}
\usepackage{dcolumn} 
\usepackage{amsmath}
\DeclareMathOperator{\erfc}{erfc} 

\renewcommand\vec[1]{\boldsymbol{#1}}

\begin{document}

\preprint{UMIST/Phys/TP/1-2}

\title{Full two-dimensional numerical study of the quantum-Hall Skyrme crystal}

\author{Niels R. Walet} \email{Niels.Walet@umist.ac.uk} \author{Tom
Weidig} 
\email{Tom.Weidig@physics.org}
\affiliation{Department of Physics, UMIST, P.O. Box 88, Manchester M60
1QD, UK}

\date{\today}

\begin{abstract}

Spin textures arise in the effective action approach to the quantum
Hall effect.  Up to now these textures, also called Skyrmions, have
been mainly studied using approximations.  Studies of its finite
density form, the crystal, have been limited to a given symmetry and
tend to ignore Skyrmion deformation. Using a simulated annealing technique,
adapted to problems with a long-range interaction, we are able to
present the first full two-dimensional study of quantum Hall Skyrmions and the
corresponding crystals. Our results show that only the Skyrmions with
topological charge one and two are bound and that the Skyrme crystal
is probably a triangular lattice made of charge-two Skyrmions.
\end{abstract}

\pacs{73.43.-f,
12.39.Dc,
05.10.-a}

\maketitle

In a strong magnetic field a two-dimensional electron gas exhibits all
the complexity of the quantum Hall effects \cite{Pekka}. The motion of
electrons is described by the Landau levels. At certain filling
factors of the lowest Landau level (notably 1 and 1/3), the ground
state becomes ferromagnetic \cite{QHFT}. The lowest excitations are no
longer single-particle states but spin textures
\cite{Skyrmetextures}, a type of solitons called Skyrmions.  The Skyrmion is a twisted two-dimensional
configuration of spins with spin down at the center and spin up far
away from the center \cite{Skyrmetextures}. A Skyrmion has a
topological charge which corresponds to the number of times the spin
rotates by $2\pi$. Its stability arises from the balance between the
electron-electron Coulomb energy (of the order of
$e^2/(4\pi\epsilon l)\approx 13.7 \text{ meV}$ and the
Zeeman energy (of the order of $g\mu_B B\approx 0.3 \text{ meV}$)
\cite{Tycko95,Barrett95}.

Since Skyrmions are the basic low-energy excitations in the quantum
Hall effect at $\nu=1$, they should play a crucial role in the
groundstate \emph{near} $\nu=1$.  It is expected that this state
contains a finite density of Skyrmions \cite{BFCD95}.  Do these
Skyrmions form a crystal? If so, what is the symmetry of the crystal?
How important are quantum fluctuations and disorder caused by
impurities? No final consensus as to the relative importance of all
these effects has yet been reached. All experimental evidence is
consistent with the existence of Skyrmions \cite{exp}. A recent paper
by Khandelwal \emph{et al.} \cite{Khan} gives evidence for Skyrmion
localization, but does not seem to fully agree with most theoretical
predictions \cite{sorry}.

One of the reasons may be that up to now the study of the Skyrme
crystal and multi-Skyrmions, Skyrmions of arbitrary topological
charge, has been performed using simplifying approximations.  Most
calculations assume the existence of a symmetry and/or start from a
simple potential picture that ignores deformation of the
Skyrmions. For multi-Skyrmions, one uses an Ansatz.  It is difficult
to numerically minimize the Skyrmion energy functional, because it
contains a non-local Coulomb term.  On the other hand, full
two-dimensional numerical studies have been done for a local version
of the two-dimensional Skyrme model, usually called the baby Skyrme
model to distinguish it from the three-dimensional one. A
thermodynamic approach was used to find its multi-Skyrmions (using
simulated annealing) \cite{HSW} and to study its thermodynamical
properties \cite{Oli2}. We have now successfully extended this
promising technique to non-local Coulomb terms.

In this Letter, we are therefore able to numerically discuss the
structure of the quantum Hall multi-Skyrmions and Skyrme crystals
without any restricting approximation.  We will also use our
experience on the thermodynamics of the local baby Skyrme model to
extrapolate to the quantum Hall model and discuss future research. We
will give more details on numerical techniques and more detailed
results in a longer paper \cite{WW}.


As mentioned, the quantum Hall ground state is ferromagnetic for
filling factors $\nu=1$ and $\nu=1/3$.  The lowest Landau level
projected quantum field theory is quite complicated and has
interesting quasi-particle states.  One can either try to use
quantum-mechanical Hartree-Fock approximations \cite{Fertig97}, or use a
semiclassical Skyrme model.  We use the latter approach, where the Skyrme
model arises through the effective action of the long-wavelength
limit of the quantum field theory. Its Lagrange density has a time
dependent term containing a gauge interaction, a gradient expansion
term, a Zeeman term, and a Coulomb interaction:
\begin{eqnarray}
\mathcal{L}_{\text{eff}}&=&\frac{1}{2}\bar{\rho} \mathcal{A}i(\partial_t\vec n)
-\frac{1}{2} \rho_s (\vec\nabla \vec n)^2 -\frac{1}{2} g \bar \rho
\mu_B \vec n \cdot \vec B \nonumber\\&&
-\nu^2_{\text{FM}}\frac{e^2}{2}\frac{1}{4\pi\epsilon} \int d^2r'\,\frac{q(\vec r)
q(\vec r')}{|\vec r-\vec r'|}\,.
\label{eq:effaction}
\end{eqnarray}
Here, $\nu_{\text{FM}}$ is the filling factor of the ferromagnetic
component of the QH state. The ferromagnetic charge density is $\bar
\rho=\frac{\nu_{\text{FM}}}{2\pi l^2}$, where the magnetic length
$l=\sqrt{\hbar /eB}$. The topological charge density, $q(\vec r)=\vec
n\cdot(\partial_x\vec n \times \partial_y\vec n)/4\pi$, describes the deviation
of the electron density from the ferromagnetic background, $\rho(\vec
r)-\bar\rho=\nu_{\text{FM}} q(\vec r)$. The spin stiffness $\rho_s=
\frac{1}{16\sqrt{2\pi}}\nu_{\text{FM}} \frac{e^2}{4\pi\epsilon l}$.
The relative dielectric constant $\epsilon_r$
($\epsilon=\epsilon_r\epsilon_0$) and the $g$ factor are properties of
the sample used.

In this Letter, we ignore the time-dependent term, because we are
looking for static solutions. Briefly, its role is to make the field
$\vec n$ satisfy the Poisson bracket $\{ \vec n_i(\vec r) , \vec
n_j(\vec r')\}=\epsilon_{ijk}\vec n_k(\vec r)\delta(\vec r-\vec r')$,
and the Hamiltonian is equal to the potential energy.

The Zeeman term is proportional to the unit of length squared and the
Coulomb term is proportional to the inverse length $d$. The gradient
term is scale invariant. The size of the static soliton solutions of
the Skyrme model is thus determined by the balance between the Coulomb
and the Zeeman term. Using a virial argument, one can show that at the
minimum energy solution the Coulomb energy must be twice the Zeeman
energy.  We find it numerically preferable to use a unit of length in
which the Zeeman and Coulomb terms balance, rather than use the
magnetic length as unit.  Equating the coefficients, we get $g \bar
\rho \mu_B B d^2= \nu_{\text{FM}}^2 e^2/(4\pi\epsilon d)$, which leads to
\begin{equation}
\frac{d}{l} = \left(  \frac{e^2 \nu_{\text{FM}} }{2 \epsilon g \mu_b B l\,.
}\right)^{1/3} 
\end{equation}
They are our units of length, and the unit of energy becomes
$e^2/(4\pi\epsilon d)$.  The energy density  reads 
($\vec x= \vec r/d$,
$\vec B=B \vec e_3$)
\begin{eqnarray}
E &=&\frac{1}{2} \Bigl( \int d^2x \left[\tilde \rho_s (\vec\nabla_x
\vec n)^2 + n_3-1 \right] \nonumber\\&& + \int d^2x\,d^2x'\,\frac{q(\vec
x) q(\vec x')}{|\vec x-\vec x'|}\Bigr) \,.\label{eq:energy}
\end{eqnarray}
This is thus effectively a one-parameter problem in terms of $\tilde
\rho_s=\frac{1}{16\sqrt{2\pi}} \nu_{\text{FM}}\frac{d}{l}$.  Note
however that the quality of the gradient expansion depends on the size
of the Skyrmions (related to $d$) as compared to the magnetic length
$l$, and thus comparison with experiment and HF calculations will only
make quantitative sense when $d \gg l$. For ``reasonable'' parameters
\begin{equation}
B=10\text{ T},\qquad 
\epsilon_r=13\text{, and}\qquad g=-0.44,
\label{eq:params}
\end{equation} these take the values
$l=8.2 \text{ nm}$ and $d/l\approx 7$.


The Skyrmions up to topological charge three have been studied in
Ref.~\cite{Norway}, using radially symmetric Ans\"atze.  In this paper
it was shown that, for a large range of parameters, the charge-two Skyrmion
is bound whereas the charge-three Skyrmion is not. They also made a
comparison with more microscopic Hartree-Fock results, and discussed
the limitations of the classical approach. Several studies of the
local baby Skyrme models show that the non-radially symmetric
multi-Skyrmions can have far lower energies than their radially
symmetric analogues. However, the exact details are highly dependent
on the form of the potential term; see \cite{WeidigBS}. Based on our
past experience, we think it is very important to do a full two-dimensional
numerical study. We first study the minimization of the energy
functional (\ref{eq:energy}). We discretize the model on a rectangular
spatial grid and impose fixed boundary conditions with the vacuum
field $\vec n=(0,0,1)$ on the boundary. We start out with a field
configuration of a given topological charge. Then we minimize the
energy functional using simulated annealing \cite{HSW}. We have
extended this minimization technique to handle the long-range interaction
terms arising from the Coulomb force.

\begin{figure}[!tbp]
\begin{center}
\includegraphics[width=4cm]{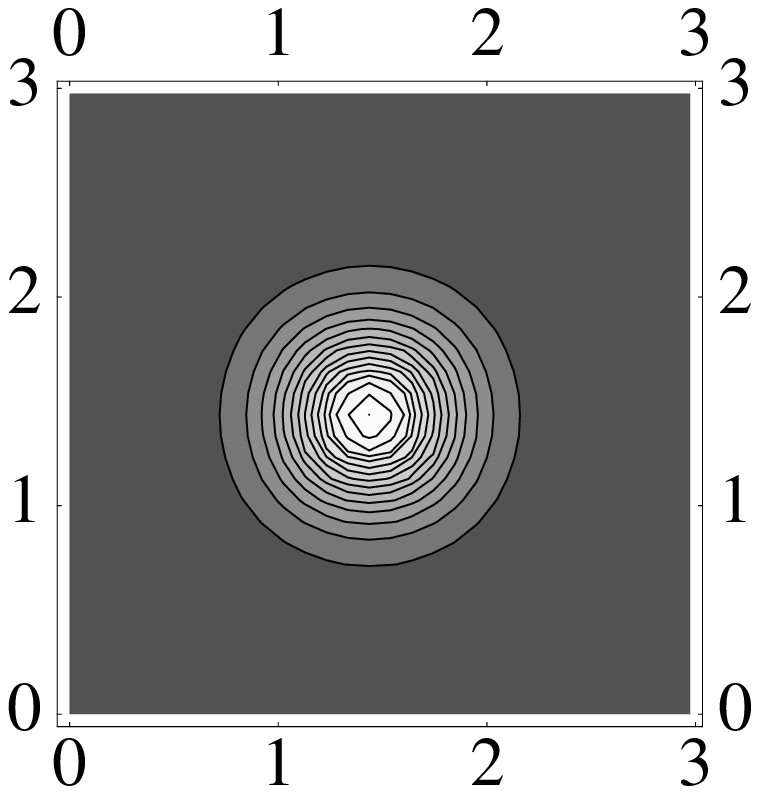}\\
\includegraphics[width=4cm]{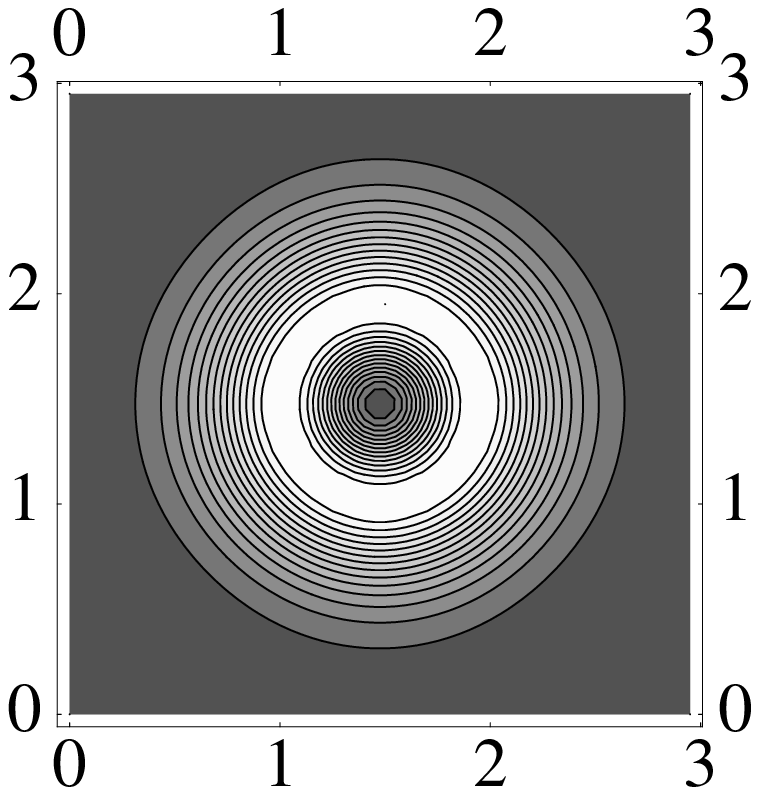}\\
\includegraphics[width=4cm]{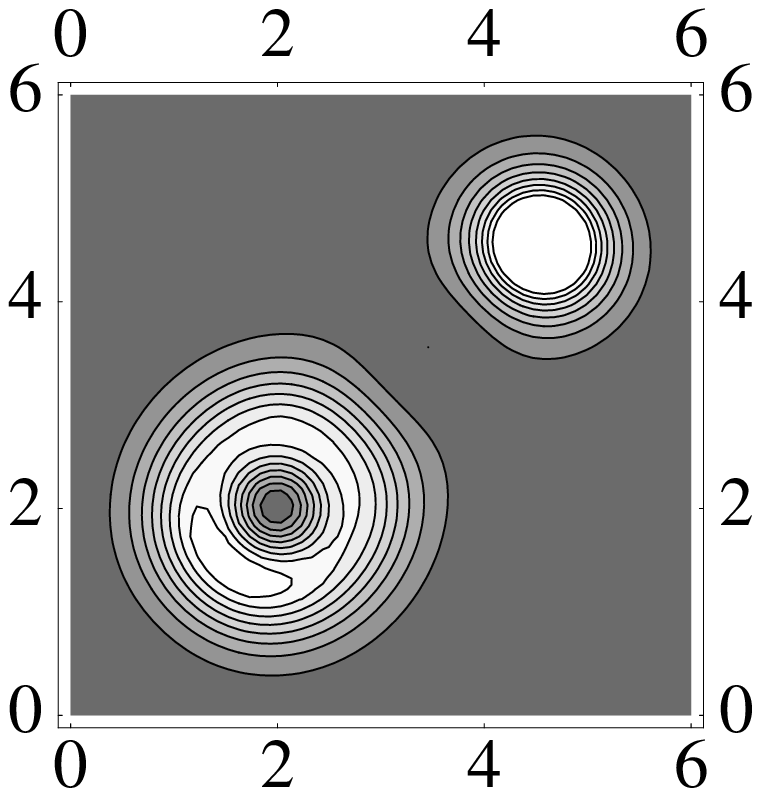}
\end{center}
\caption{The Skyrmions from topological charge one to three, for the
parameter $d/l=7$. Lighter colors denote a higher charge density, and
lengths are expressed in units of $d$.}
\label{fig:multisol}.
\end{figure} 

\begin{figure*}
\centerline{\includegraphics[width=3.8cm]{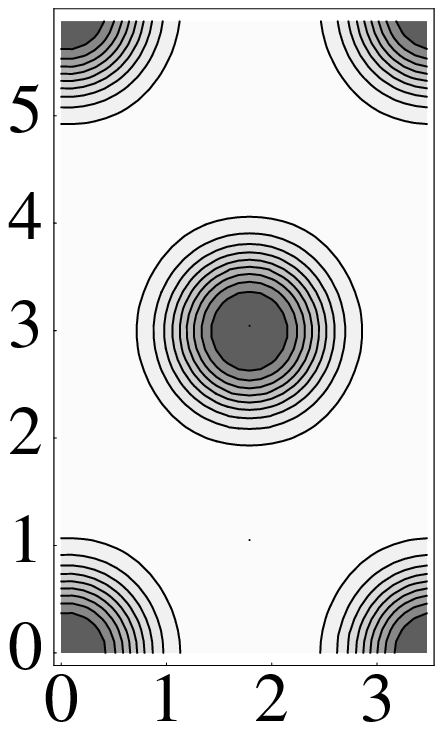}\hspace*{0.5cm}
\includegraphics[width=3.8cm]{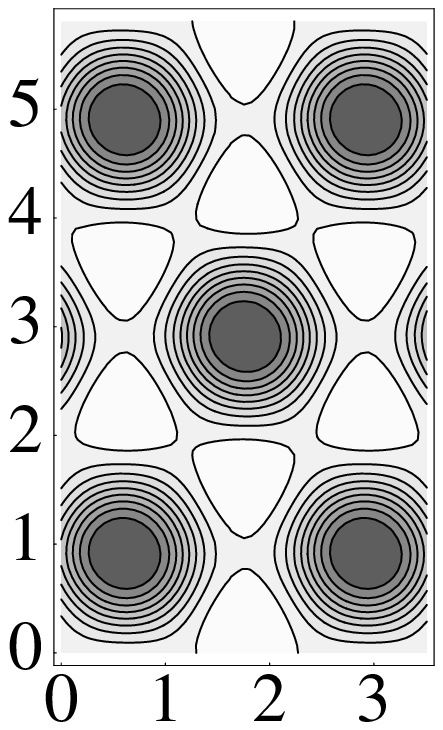}\hspace*{0.5cm}
\includegraphics[width=3.8cm]{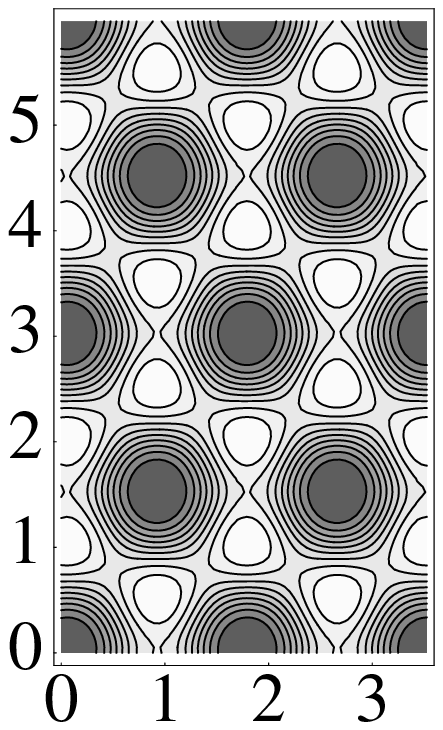}}
\caption{Magnetisation in triangular charge-two Skyrmion crystals, 
for increasing densities.
The units of length is given in terms of the parameter $d$; for $d/l=7$ 
the change in filling of the quantum-Hall state is (from left to right)
$\Delta\nu=0.024$, $0.0544$, and $0.0967$. Light colors denote a spin pointing
upwards; the darkest denote a downwards spin. The contour levels are the same
across all three graphs.
See also \cite{Figs}.}
\label{fig:crystal}.
\end{figure*}

We have investigated the minimal energy solutions with the topological
charge one to four; see Fig.~\ref{fig:multisol} for some results.
We found that only the charge-two Skyrmion is bound. The simulations
for charge three and four seem to reveal a bound charge-three Skyrmion
and a isosceles-triangular charge-four Skyrmion.  However, these are
finite box effects. When increasing the box size we are left with
individual Skyrmions of charge one and two. As energetically expected,
a charge-three field splits into a charge-one and a charge-two
Skyrmion, and the charge-four Skyrmion splits into two charge-two
Skyrmions. We see clear indication that the charge-two Skyrmion might
be the building block of any quantum Hall Skyrme crystal. It is very
likely that a change in the parameters of the Lagrangian,
and modification of the potential are able to
dramatically change the structure of the multi-Skyrmions and possibly
lead to stable Skyrmions above charge two. In fact, any change of the
effective action, such as higher-order gradient expansion terms,
medium modifications to the Coulomb potential (see below), 
quantum effects, and inhomogeneities can have an impact on weakly 
bound multi-Skyrmions.

We now turn our attention to the study of the quantum Hall Skyrme
crystal, and produce a full two-dimensional study. We do not impose
any symmetry, apart from that imposed by the periodic boundary
conditions of the lattice.  We need to be very careful when studying
crystals with non-local Coulomb interaction terms.  The crystal could
be thought of as densely packed multi-Skyrmions, the quasi-particles
of the Skyrme model, but the Skyrmions are all charged, and the
samples used to study Skyrmions are electrically neutral.  This means
that Skyrme crystals exist in a positively charged background.  This
also renormalizes away the infinite Coulomb energy otherwise carried
by a crystal.  Since we shall use a finite size box for the
simulation, we can use an Ewald sum rule to evaluate the sum over all
the image charges generated by the periodic boundary conditions.  This
method explicitly gives us the zero-wave length component of the
Coulomb field which we then renormalize away (a more detailed analysis
of the procedure will be given in Ref.~\cite{WW}).  The first two
parts of the energy expression (\ref{eq:coulomb}) are not modified
when applying periodic BCs, but the Coulomb part becomes a sum over
the vectors $\vec R=(n_1 L_1, n_2 L_2)$, where $L_{1,2}$ are the
lengths of the simulation volume,
\begin{equation}\label{eq:coulomb}
\frac{1}{|\vec x - \vec x'|} \rightarrow \sum_{\vec R} \frac{1}{|\vec
x - (\vec x'+\vec R)|}\,.
\end{equation}
Using Ewald's technique, we split the sum
into a sum over the real and the reciprocal lattice (the integers $k$ and
$l$  are used in the sum over the real lattice, $k'$ and $l'$ for the
reciprocal one)
\begin{align}
V&(s_{x},s_{y}) \sum _{\vec{R}}\frac{1}{|\vec{s}+\vec{R}|}\nonumber\\ & = 
 \sqrt{1/(L_{1}L_{2})}
\biggl\{\sum
 _{k'l'}e^{i2\pi (k' s_x/L_1+l's_y/L_2)}\nonumber\\&\qquad\times
 \Phi \left( \pi \left[ {k'}^{2}L_{2}/L_{1}+{l'}^{2}L_{1}/L_{2}\right]
 \right) +\nonumber\\ &   +\sum _{kl}\Phi \left( \pi \left[
 (k+\frac{s_x}{L_1})^{2}\frac{L_{1}}{L_{2}}+(l+\frac{s_y}{L_2})^{2}\frac{L_{2}}{L_{1}}\right]
 \right)  \biggr\} ,
\label{eq:Ewald}
\end{align}
where
\[
\Phi (x)=\frac{\sqrt{\pi }}{\sqrt{x}}\erfc (\sqrt{x})\,.\] The term
$k'=l'=0$ is infinite. 
Adding a positive background charge leaves us a constant energy shift 
\cite{WW}.  Using (\ref{eq:Ewald}) minus this shift is simple and
straightforward. The sums all converge extremely rapidly.

\nopagebreak
We have performed a study of these finite density quantum Hall
systems, for the same parameters as the multi-Skyrmion simulations.
Of course, we are already biased by our study of the multi-Skyrmions
and expect a lattice of deformed charge-two Skyrmions. Our best guess
is a triangular lattice, because it is the preferred lattice of the
local baby Skyrme model \cite{Oli2}. Our goal was two-fold: to find the
energetically favored symmetry of the lattice of charge-two Skyrmions
and to show that other lattice construction are not stable. We started
out with many different configurations, mostly lattices made out of
charge-one Skyrmions. Nearly all configuration minimized to a
configuration made out of charge-two Skyrmions. Taking finite box
effects into account, the lattice seems to be triangular. Indeed we
believe that the energetically preferred crystal is the triangular
lattice made out of deformed charge-two Skyrmions; see
Fig.~\ref{fig:crystal}. There appears to be a rather deep connection
between these results and the complex (CP1) parametrization of the
Skyrme field in terms of Weierstrass functions, which will not be
pursued here, see however
\cite{WW}.

The thermodynamics of the quantum Hall Skyrme model is another
interesting challenge. What are the phases of the model at finite
density and temperature? Our simulated annealing already contains the
thermodynamic partition function, so it is well suited to tackle this
question: see the phase portrait of the baby Skyrme
model \cite{Oli2}. However, the model differs in several important
aspects. The baby Skyrme model has a local fourth-order term rather
than the global Coulomb term. The fact that the dynamics of the Skyrme
field is very different may have an influence on the thermodynamics as
well.  There  other kinetic terms have also been discussed,
see Ref.~\cite{Stoof}. Nonetheless, we expect some similarities,
because the crystals are similar.

The most thorough discussion of the phase diagram of Skyrmion crystals
based on a ``simple'' potential Ansatz is given in
Ref.~\cite{TGF98}. Many interesting phases are discussed, and we have
tried to simulate a few of those, by starting a simulation in a
periodic box with a Skyrme field of the right symmetry. In all cases
we found a rapid decay to crystals of charge-two Skyrmions, or extremely
frustrated crystals (due to the small box used). This does not mean
that such crystals cannot occur, but it seems to be a generic feature
of Skyrme-like theories to have a toroidal charge-two Skyrmion that is
bound: It seems to be this feature that is driving our results.  Thus
we would like to argue that at sufficiently high densities the Skyrme
crystal looks like the ones shown in Fig.~\ref{fig:crystal}. At lower densities the
behavior is different. As long as we stick with the effective action
(\ref{eq:effaction}) we shall see isolated clusters with charge two; if
we modify the interaction, we expect connected networks, or even
random clusters of multi-Skyrmions, as discussed in
Ref.~\cite{Oli2}. We intend to investigate these questions further.

The occurrence of a fluid at finite temperatures, which might well be
the phase transition seen in experiment, is also of relevance. It has
been argued that quantum fluctuations play a crucial role here as well
\cite{BP99}, so the question about the nature of the fluid and the
phase transition deserves further study.

The authors would like to thank Dr.\ O Schwindt, Dr.\ P.
Sutcliffe and Prof.\ N. Manton for useful discussions. This work
was supported by research grants (GR/L22331 and GR/N15672) from the
Engineering and Physical Sciences Research Council (EPSRC) of Great
Britain. T.W. Would like to thank Richard Battye and Trinity college
for support during a stay at DAMTP.

\end{document}